\documentclass[review]{elsarticle}

\usepackage{lineno,hyperref}
\usepackage{xspace}
\modulolinenumbers[3]
\usepackage{bm}% bold math
\usepackage{amssymb}
\usepackage{hyperref}
\usepackage{amsmath}
\usepackage{graphicx}
\usepackage[utf8]{inputenc}

\newcommand{\pt}{\ensuremath{p_{\rm{T}}}\xspace}
\newcommand{\ptjet}{\ensuremath{{p_{\rm{T}}^{\rm jet}}}\xspace}

\journal{Journal of \LaTeX\ Templates}

\begin{document}

\begin{frontmatter}

\title{Probing color reconnection with underlying-event observables at the LHC energies}

%% Group authors per affiliation:
\author{Antonio Ortiz}
\address{Instituto de Ciencias Nucleares, Universidad Nacional Aut\'onoma de M\'exico, \\ 
Ciudad de M\'exico, Del. Coyoac\'an 04510, M\'exico}

\cortext[mycorrespondingauthor]{Corresponding author}
%\ead{antonio.ortiz@nucleares.unam.mx}

\author{Lizardo Valencia Palomo}
\address{Departamento de Investigac\'ion en F\'isica, Universidad de Sonora, \\ Blvd Luis Encinas y Rosales S/N, Col. Centro, Hermosillo, Sonora, M\'exico}

%\cortext[mycorrespondingauthor]{Corresponding author}
\ead{lizardo.valencia@unison.mx}
%\ead{xxx@cern.ch}

\begin{abstract}

In this work we study the underlying-event (UE) activity as a function of the highest jet or charged-particle transverse momentum ($p_{\rm T}^{\rm jet}$ or $p_{\rm T}^{\rm leading}$) in terms of number and summed-$p_{\rm T}$ densities of charged particles in the azimuthal region transverse to the leading jet or particle direction. Contrary to inelastic $pp$ LHC data, the UE observables normalised to the charged-particle density obey an approximate Koba-Nielsen-Olesen (KNO) scaling. Based on PYTHIA~8.2 simulations of $pp$ collisions at LHC energies, we show that the small breaking of the KNO scaling is due to the increasing importance of multiple partonic interactions (MPI) at higher $\sqrt{s}$. This in turn makes that with increasing energy, the $p_{\rm T}$ spectra in the UE get harder than in inelastic $pp$ collisions.

Color reconnection (CR) models the interactions among outgoing partons just before the hadronization, therefore it modifies the $p_{\rm T}$-spectral shape. Motivated by this, we studied the UE activity considering charged particles within different $p_{\rm T}$ intervals in $pp$ collisions at $\sqrt{s}=7$\,TeV. Although MPI saturate for $p_{\rm T}^{\rm jet}>10$\,GeV/$c$, the UE still  increases with increasing $p_{\rm T}^{\rm jet}$.  We demonstrate that the saturation of both number and summed-$p_{\rm T}$ densities, which are commonly claimed, are only observed for low-\pt charged particles ($0.5<\pt<2$\,GeV/$c$).   Moreover, for the $p_{\rm T}$-integrated case ($\pt>0.5$\,GeV/$c$) the summed-$p_{\rm T}$ density is not sensitive to the variation of CR, however at low-\pt it is reduced with increasing CR, whereas an opposite behavior is found at intermediate-$p_{\rm T}$ ($2<\pt<10$\,GeV/$c$). Finally, we show that CR produces flow-like behavior only in the UE region and the effects are reduced with increasing  $p_{\rm T}^{\rm jet}$ due to the hardening of UE.  The outcomes encourage the measurement of inclusive and identified charged-particle $p_{\rm T}$ spectra (over a wide range of $p_{\rm T}$) associated to UE aimed at better understanding the similarities between $pp$ and heavy-ion data discovered at the LHC.

\end{abstract}


\end{frontmatter}

\linenumbers

\section{Introduction}

In the context of event generators, one single inelastic (non-diffractive) proton-proton ($pp$) collision can be split into two components: the main hard partonic scattering and the underlying event (UE). The latter includes initial- and final-state radiation (ISR/FSR), multiple partonic interactions (MPI) and the fragmentation of beam remnants resulting from the hadronization of the partonic constituents that did not participate in any scatter. On top of that, color reconnection (CR) is a microscopic mechanism that describes the interactions that can occur between color fields during the hadronization transition.   It is a key ingredient which is needed to explain the increase of the average transverse momentum as a function of charged-particle multiplicity~\cite{PhysRevD.36.2019} and the multiplicity-dependent event shape studies~\cite{Abelev:2012sk}. The importance of the UE relies on the fact that it can shed light in the search for new physics phenomena at hadron colliders~\cite{Martin:2016igp} or even in the quest for precise Standard Model measurements~\cite{Harnik:2008ax}.  

The UE observables have been measured in $p\bar{p}$ collisions in dijet and Drell-Yan processes at CDF at $\sqrt{s}=1.8$ and 1.96 TeV~\cite{Acosta:2004wqa,Aaltonen:2010rm}. At the start of the LHC most of the existing Monte Carlo models were tuned using Tevatron data. In this sense, early measurements provided by the ATLAS experiment~\cite{Aad:2010fh} showed clear differences between event generators and data. Latest results have shown an improvement on the description of the UE by the new Monte Carlo tunes~\cite{Aaboud:2017fwp}. However, we should keep in mind that the modelling of UE is more challenging than traditionally assumed because recent results suggest the unexpected presence of heavy ion-like behavior in $pp$ data~\cite{Ortiz:2015cma,Abelev:2013haa,Khachatryan:2016txc,ALICE:2017jyt}. We should therefore go beyond the traditional UE analysis for better understanding the similarities between $pp$ and heavy-ion data. 

In order to understand the physics behind high-multiplicity $pp$ collisions, we have performed several studies using PYTHIA~8~\cite{Sjostrand:2014zea} event generator. We have found that CR together with MPI can produce radial flow patterns via boosted color strings~\cite{Ortiz:2013yxa}. The analysis as a function of event multiplicity allows to study events with different amount of MPI, however we have shown that high-multiplicity $pp$ collisions are effected by the so-called fragmentation bias~\cite{Abelev:2013sqa,Ortiz:2016kpz}. Moreover, for a fixed amount of MPI CR strongly modifies the charged-particle multiplicity, therefore a selection based on multiplicity does not permit the isolation of CR effects in events with the same MPI activity. In this work we propose to exploit the usage of  UE in order to increase the sensitivity to events with multiple semi-hard scatterings instead to biased fragmentation of high transverse momentum jets. Similar ideas have been proposed in order to disentangle auto-correlation effects from other physical phenomena by measuring the charged-particle multiplicity in the UE region~\cite{Weber:2018ddv}.  

Given that heavy-ion data feature remarkable differences between the particle production associated to bulk (everything outside the jet peaks) and jets~\cite{Veldhoen:2012ge,Richert:2016bwd}, we have decided to perform an analogous analysis in $pp$ collisions simulated with PYTHIA~8. The advantage is that cutting on the highest charged-particle transverse momentum ($\pt^{\rm leading}$) or highest jet \pt (\ptjet) it is possible to select events with larger (and nearly constant) MPI activity than that associated to inelastic events. This allows to enhance color-reconnection effects on the hadronization. 

Specifically, in this paper we study the modification of the traditional UE observables when charged particles within different \pt intervals are considered. We also aim to understand how the underlying-event activity is affected when either \ptjet or $\sqrt{s}$ are varied. The analysis is not restricted to the traditional average quantities, instead we study the \pt spectra of unidentified and identified charged particles within UE environments modified with a selection on \ptjet. The measurements of the \pt spectra sensitive to UE as a function of \ptjet are proposed as a tool to probe the QCD Monte Carlo picture using LHC data.  

The article is organised as follows: section 2 provides information about the event and particle selection for the simulations, section 3 discusses the energy dependence of the UE in terms of the number and summed-\pt densities, section 4 presents the results on how CR affects the UE observables and the \pt distribution of identified charged particles, and finally section 5 contains the summary and outlook.

\section{Simulations and analysis}

Quantities sensitive to the UE are constructed as follows. The perpendicular plane to the beam axis is segmented into three regions depending on the azimuthal angular difference ($\Delta \phi$) relative to the leading object (e.g. jet or charged particle) axis. Unless specified, those particles with $\pi/3 < \lvert \Delta \phi \rvert < 2\pi/3$  (transverse side) are the only ones used for the present analysis since they are the most sensitive to UE. The underlying-event observables are defined as the mean number of charged particles per pseudorapidity unit and per azimuthal separation unit ($\langle N_{\mathrm{ch}}/\Delta\eta\Delta\phi \rangle$) and the mean scalar sum of transverse momentum per pseudorapidity unit and per azimuthal separation unit ($\langle \Sigma p_{\mathrm{T}}/\Delta\eta\Delta\phi \rangle$). hereafter $\langle N_{\mathrm{ch}}/\Delta\eta\Delta\phi \rangle$ and $\langle \Sigma p_{\mathrm{T}}/\Delta\eta\Delta\phi \rangle$ are referred to as number density and summed-\pt density, respectively. PYTHIA~8.212~\cite{Sjostrand:2014zea} tune Monash 2013~\cite{Skands:2014pea}, hereafter referred to as PY\-THIA~8, was used to simulate inelastic $pp$ collisions at the LHC energies. Since simulations are intended to be compared to CMS~\cite{Chatrchyan:2011id}, ATLAS~\cite{Aaboud:2017fwp} and ALICE~\cite{ALICE:2011ac} results available on HEPData~\cite{Maguire:2017ypu}, different selections at particle and event levels were implemented:

\begin{enumerate}
\item The comparison of simulations with CMS data ($pp$ collisions at $\sqrt{s}=0.9$ and 7\,TeV) requires a leading jet with transverse momentum larger than 1\,GeV/$c$ within the pseudorapidity interval: $\lvert\eta\rvert < 2$. Jets are defined using the SISCone algorithm~\cite{Salam:2007xv} as implemented in FastJet 3.3~\cite{Cacciari:2011ma} with the clustering radius given as $R = \sqrt{(\Delta \phi)^2 + (\Delta \eta)^2} = 0.5$. Furthermore, jets are built using charged particles with $\pt> 0.5$\,GeV/$c$ and $\lvert\eta\rvert < 2.5$.  The UE observables are computed using charged particles within $\lvert\eta\rvert < 2$ and $\pt>0.5$\,GeV/$c$. Notice the $\eta$ range of particles used to build the jets is larger than that of particles in the UE in order to avoid a kinematic bias.

\item The comparison with ALICE data ($pp$ collisions at $\sqrt{s}=0.9$ and 7\,TeV) requires charged particles with $\pt>0.5$ GeV/$c$ and $|\eta|<0.8$ with the transverse momentum of the leading charged particle above 0.5\,GeV/$c$.

\item The comparison with ATLAS data ($pp$ collisions at $\sqrt{s}=13$\,TeV) requires charged particles with $\pt>0.5$ GeV/$c$ and $|\eta|<2.5$ with the transverse momentum of the leading charged particle above 1\,GeV/$c$.

\end{enumerate}

In all cases the configuration used in PYTHIA~8 was the following: Beams:idA = 2212, Beams:idB = 2212, Beams:eCM = 900 (7000 and 13000), Tune:pp = 14 and SoftQCD:inelastic = on.

\section{Energy dependence of the underlying event}

In a previous publication~\cite{Ortiz:2017jaz}, we have shown that within 10\% the UE observables as a function of $p_{\rm T}^{\rm leading}$ collapse on $\sqrt{s}$-independent curves once they are scaled to the corresponding average charged-particle density (${\rm d}N_{\rm ch}/{\rm d}\eta^{\rm inelastic}$). This quantity is obtained for inelastic $pp$ collisions at a given center-of-mass energy. In this paper we want to trace back the origin of the small imperfection of the scaling behavior, i.e. its validity within 10\%. Instead of using ATLAS UE data (measured as a function of $p_{\rm T}^{\rm leading}$), now we use the CMS data since they are reported for a wide \ptjet interval. This allows the study of $pp$ events with exceptional high-\pt jets.  In order to obtain the charged-particle density, the transverse momentum distributions of charged particles in $pp$ collisions at $\sqrt{s}=0.9$ and 7\,TeV~\cite{Khachatryan:2010xs,Khachatryan:2010us} were integrated for $\lvert\eta\rvert < 2.4$ and $\pt> 0.5$\,GeV/$c$. The total uncertainties (quadratic sum of statistical and systematic uncertainties) associated to the \pt spectrum were propagated to the integral and assigned as systematic uncertainty to $\langle {\rm d}N_{\rm ch}/{\rm d}\eta \rangle^{\rm inelastic}$. For $pp$ collisions at $\sqrt{s} = 0.9$ and 7 TeV, the computed values were $\langle {\rm d}N_{\rm ch}/{\rm d}\eta\rangle^{\rm inelastic} = 1.20 \pm 0.04$ and $2.33 \pm 0.11$, respectively.  Therefore, the relative variation ($f$) of the average charged-particle density in $pp$ collisions at $\sqrt{s}=7$\,TeV with respect to $pp$ collisions at $\sqrt{s}=0.9$ TeV was found to be $f = 1.94 \pm 0.11$.

\begin{figure*}
\begin{center}
\includegraphics[width=0.49\textwidth]{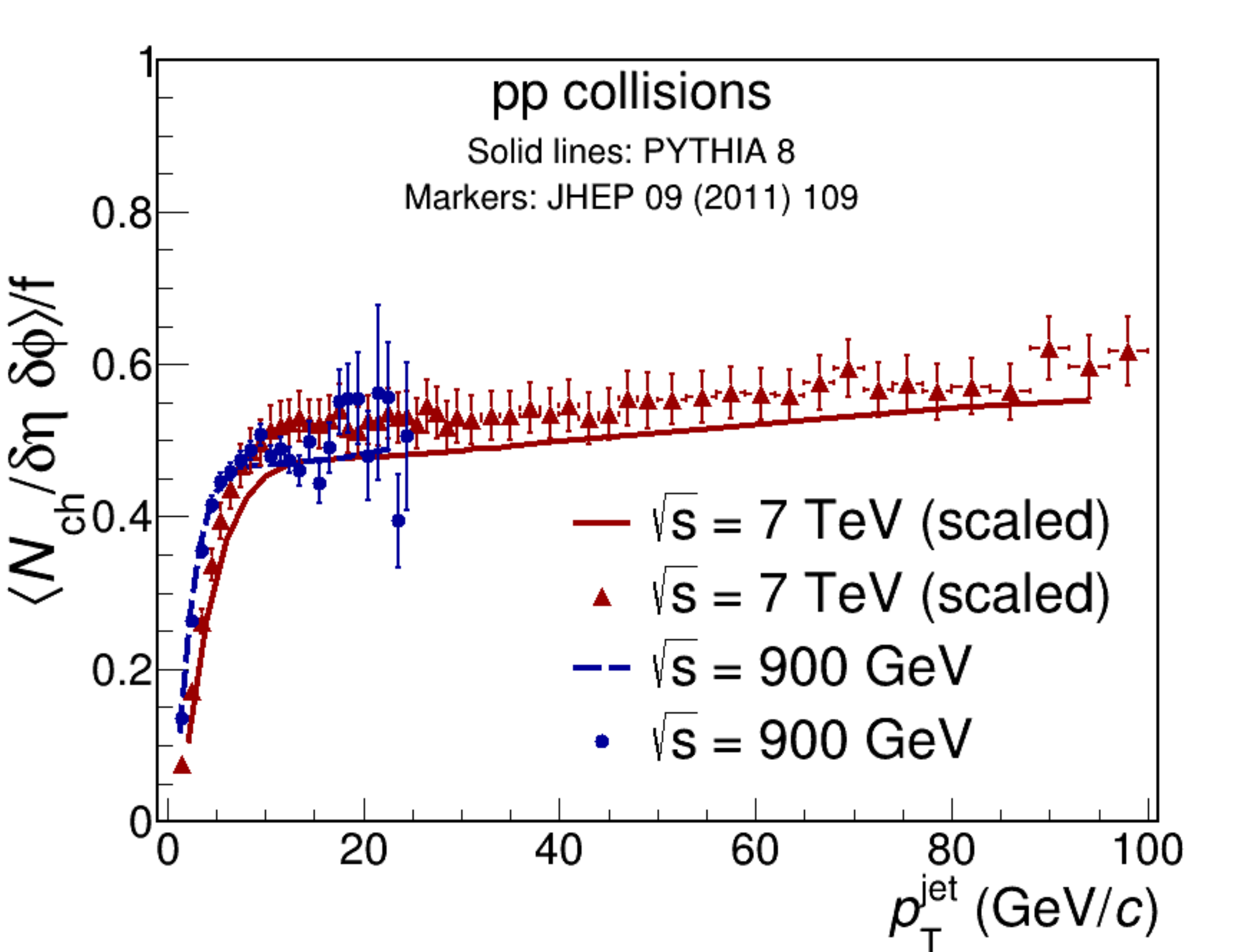}
\hspace{-0.1cm}
\includegraphics[width=0.49\textwidth]{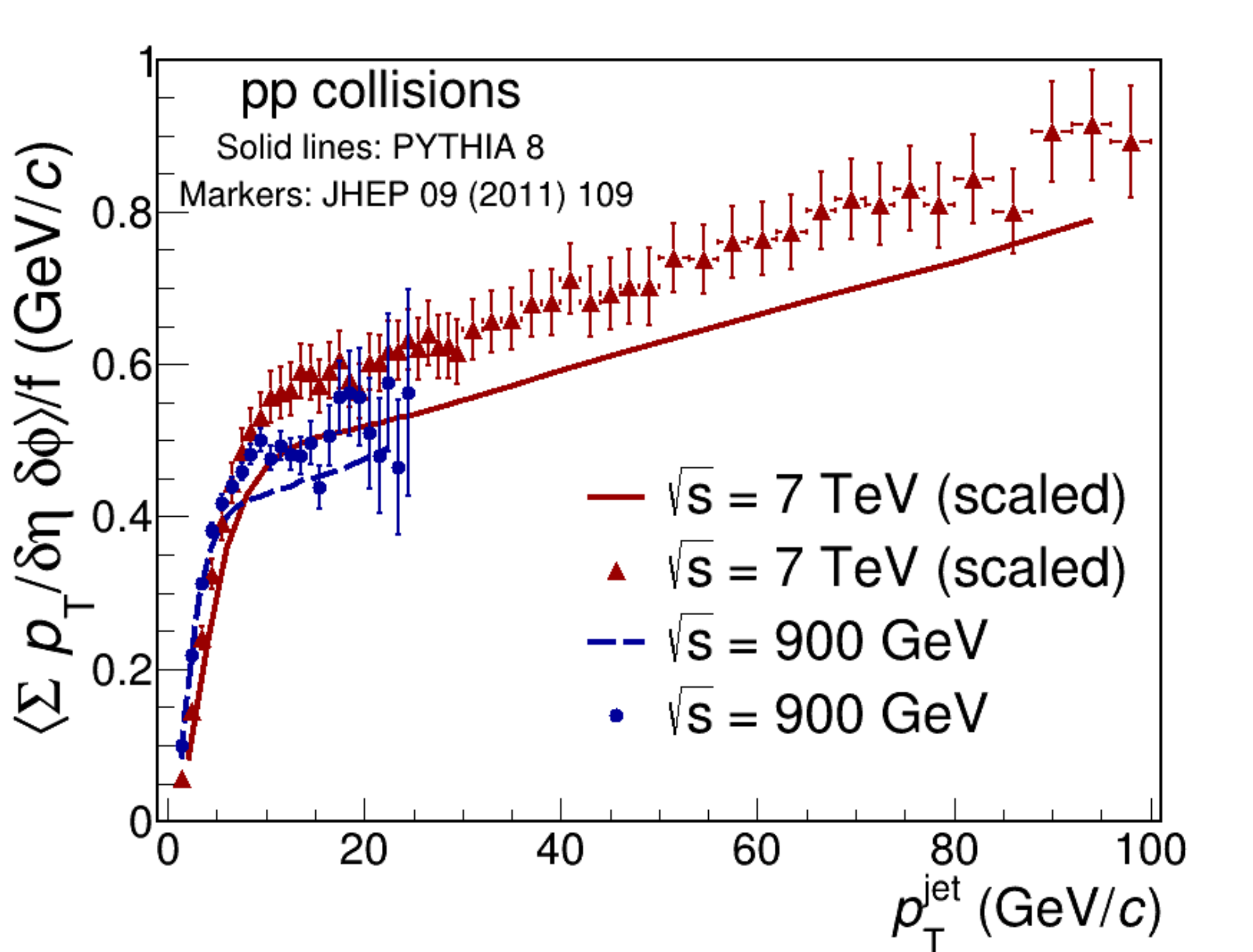}
\caption{ (Color online). The number density (left) and the summed-\pt density (right) as a function of \ptjet in the UE region. A factor $f$ that takes into account the relative charged-particle density increase from 0.9 to 7 TeV has been applied. Data are shown as markers and the error bars represent the total uncertainties (the one measured by CMS and the associated to $f$). Solid lines are the results obtained with PYTHIA~8.}
\label{Scaling}
\end{center}
\end{figure*}

\begin{figure*}
\begin{center}
\includegraphics[width=0.49\textwidth]{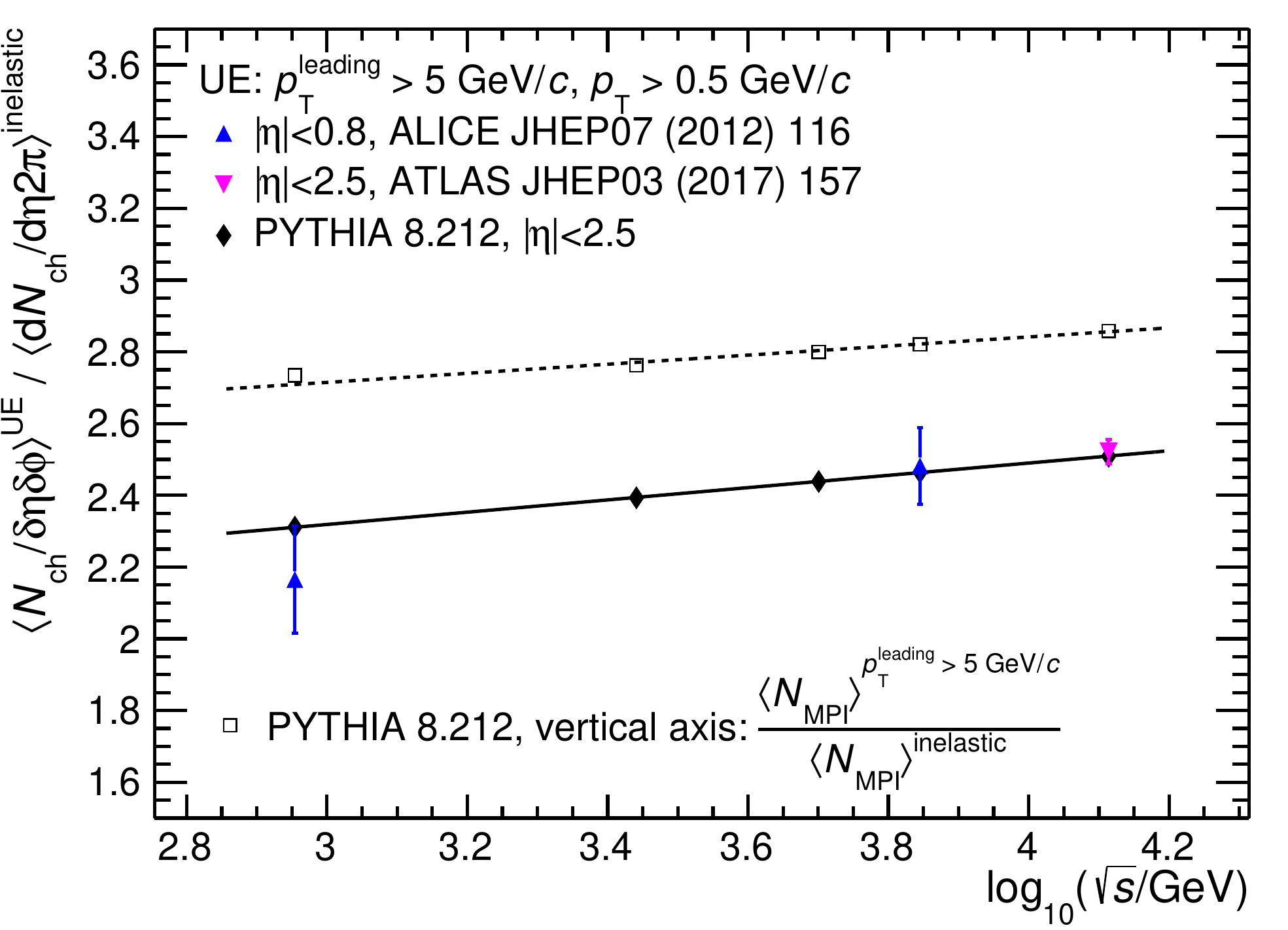}
\hspace{-0.1cm}
\includegraphics[width=0.49\textwidth]{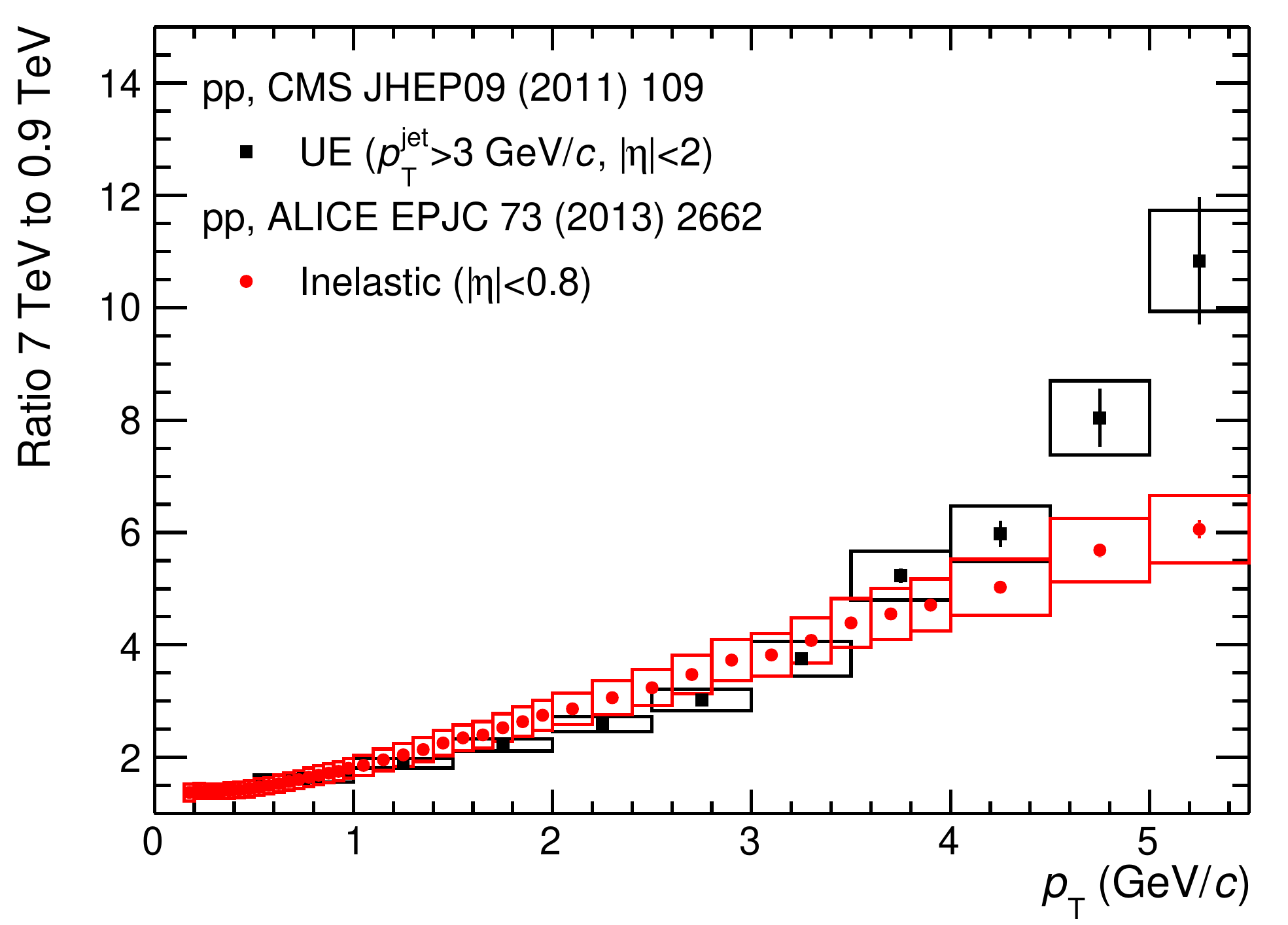}
\caption{ (Color online). Left: Energy dependence of the number density (UE) normalised to the average charged-particle density in inelastic events. PYTHIA 8 simulations of $pp$ collisions at $\sqrt{s}=0.9$, 2.76, 5.02, 7 and 13 TeV (full black markers) are compared to ALICE (full blue markers) and ATLAS data (full magenta markers). Similar ratios (empty markers) were obtained using the number of multiple partonic interactions ($N_{\rm MPI}$). Right: UE transverse momentum distribution (associated to $\ptjet>3$\,GeV/$c$) in $pp$ collisions at 7\,TeV normalized to the corresponding one at 0.9\,TeV (black squares). Results are compared to analogous ratios for inelastic $pp$ collisions at the same energies (red circles). Statistical and systematic uncertainties are represented as error bars and boxes around the data points, respectively. Data were obtained from~\cite{Chatrchyan:2011id,Abelev:2013ala}.}
\label{Scaling2}
\end{center}
\end{figure*}

Figure~\ref{Scaling} shows the number (left panel) and the summed-\pt (right panel) densities as a function of \ptjet after scaling by a 1/$f$ factor. Little or no energy dependence is observed after the implementation of the scaling factor. This is what we called ``universality of UE'' in our previous work where $\pt^{\rm leading}$ was used instead~\cite{Ortiz:2017jaz}. The results are compared with PYTHIA~8 simulations since it reproduces very well many features of LHC data. We observe that the scaling properties hold up to $p_{\rm T}^{\rm jet}=25$\,GeV/$c$. Regarding the behavior of data, a steep rise of the underlying activity with increasing $p_{\rm T}^{\rm jet}$ is observed. This fast rise is followed by a change of the slope above $p_{\rm T}^{\rm jet} \sim 10$\,GeV/$c$. The change of the slope is understood as due to very small impact parameter $pp$ collisions which yields to a saturation of the MPI activity.  Different publications state that above 10\,GeV/$c$ the UE observables saturate, reaching nearly constant values~\cite{ALICE:2011ac,Aaboud:2017fwp,Khachatryan:2015jza}. However, we want to stress the fact that both data and PYTHIA~8 still show a rise of the UE activity with increasing \ptjet. In the next sections we discuss how color reconnection contributes to that behavior.  

The universality of UE can be interpreted as the approximate Koba-Nielsen-Olesen (KNO)~\cite{Koba:1972ng} scaling of the particle production associated to the UE.  For this reason we studied the energy dependence of the KNO variable: number density in the transverse side normalised to the charged-particle density (inelastic $pp$ collisions). This ratio has been studied by ALICE using $pp$ data at $\sqrt{s}=0.9$ and 7\,TeV~\cite{ALICE:2011ac}. To obtain the numerator, ALICE fitted a first degree polynomial to the number density as a function of $p_{\rm T}^{\rm leading}$ for $p_{\rm T}^{\rm leading}>5$\,GeV/$c$. Within uncertainties the fit slope was found to be consistent with zero and the constant parameter was identified as the average number density in the transverse side. Potential correlations of the systematic uncertainties in different \pt bins were neglected. The left-hand side of Fig.~\ref{Scaling2} shows a comparison of the KNO variable as a function of $\log(\sqrt{s})$ measured by ALICE and ATLAS. ALICE data points were taken from~\cite{ALICE:2011ac} while the ATLAS point was derived following the strategy already described. PYTHIA~8 inelastic $pp$ events were simulated in order to test the scaling of UE over the wide $\sqrt{s}$ interval covered at the LHC. Going from $pp$ collisions at $\sqrt{s}=0.9$ to 13\,TeV, PYTHIA~8 simulations show a $\sim8\%$ increase of the KNO variable. Within uncertainties, the data exhibit the same trend. The small $\sqrt{s}$ dependence is also observed if we consider the average number of MPI in events with $p_{\rm T}^{\rm leading}>5$\,GeV/$c$ (central $pp$ collisions) normalised to this same observable but in inelastic $pp$ collisions (i.e. without any selection on $p_{\rm T}^{\rm leading}$). The ratio as a function of $\log({\sqrt{s}})$ is described by a first degree polynomial with a slope parameter of 0.13, which is close to what is obtained using the number density (slope 0.17), suggesting a common origin.

In order to investigate the features of particle production associated to UE, in particular the excess of particles in UE with respect to inelastic $pp$ collisions, we studied the \pt spectra of unidentified charged particles. The CMS collaboration has reported such spectra for $pp$ collisions at $\sqrt{s}=0.9$ and 7\,TeV. The UE corresponds to events with $\ptjet>3$\,GeV/$c$ within $|\eta|<2$. To study the evolution of the \pt-differential particle production with increasing energy, the 7\,TeV \pt spectrum was normalized to the corresponding one at 0.9\,TeV. The results are presented in the right-hand side of Fig.~\ref{Scaling2}.  An analogous ratio as a function of \pt was obtained using the \pt spectra of inelastic $pp$ collisions at the same energies~\cite{Abelev:2013ala}. Within uncertainties, both ratios are consistent up to $\pt=4$\,GeV/$c$. For higher \pt the ratio increases faster for the UE activity than for inelastic $pp$ collisions. The results suggest that, with increasing $\sqrt{s}$, intermediate-\pt particle production increases faster in the UE than in inelastic $pp$ collisions. This effect could explain the faster rise of the number density in the UE relative to the corresponding inelastic events~\cite{ALICE:2011ac}. Transverse momentum distributions sensitive to UE for other $\sqrt{s}$ and \ptjet are needed in order to confirm the validity of this conclusion. 

\section{CR studies using UE observables}

We have shown that, with increasing $\sqrt{s}$, the \pt spectra sensitive to UE get harder than that in inelastic events as a consequence of the $\sqrt{s}$ dependence of MPI.  Since in events with large number of MPI color reconnection modifies the particle production~\cite{Ortiz:2013yxa}, CR effects on UE are expected. In order to quantify such effects we used the default CR model of PYTHIA~8  (MPI-based) . The model includes the reconnection probability  ($rr$) to join low and high \pt partons. Therefore, $rr$ (ColourReconnection:range in PYTHIA~8) allows to control the amount of interactions which partons feel just before the hadronization. For our study, we computed $\langle N_{\mathrm{ch}}/\Delta\eta\Delta\phi \rangle$ and $\langle \Sigma p_{\mathrm{T}}/\Delta\eta\Delta\phi \rangle$ in three different \pt ranges depending on the transverse momentum of the charged particles present in the UE: integrated ($p_{\mathrm{T}} > 0.5$\,GeV/$c$), low ($0.5 < p_{\mathrm{T}} < 2$\,GeV/$c$) and intermediate ($2 < p_{\mathrm{T}} < 10$\,GeV/$c$) transverse momentum. Each \pt-dependent UE quantity was in turn computed for three different values of $rr$: 0, 1.8 and 10, corresponding to the minimum, nominal (tuned to LHC data~\cite{Skands:2014pea}) and maximum possible values in PYTHIA~8.   Moreover, the results presented here are obtained using leading jets within the CMS kinematic ranges, allowing the comparison between data and PYTHIA~8 over a broad \ptjet interval. Since we claim an approximate scaling of the particle production associated to UE, we report results only for $pp$ collisions at $\sqrt{s}=7$\,TeV. The conclusions should hold for other energies.

\paragraph{Modification of the particle production at low and intermediate \pt}

\begin{figure*}
\begin{center}
\includegraphics[keepaspectratio, width=1.0\columnwidth]{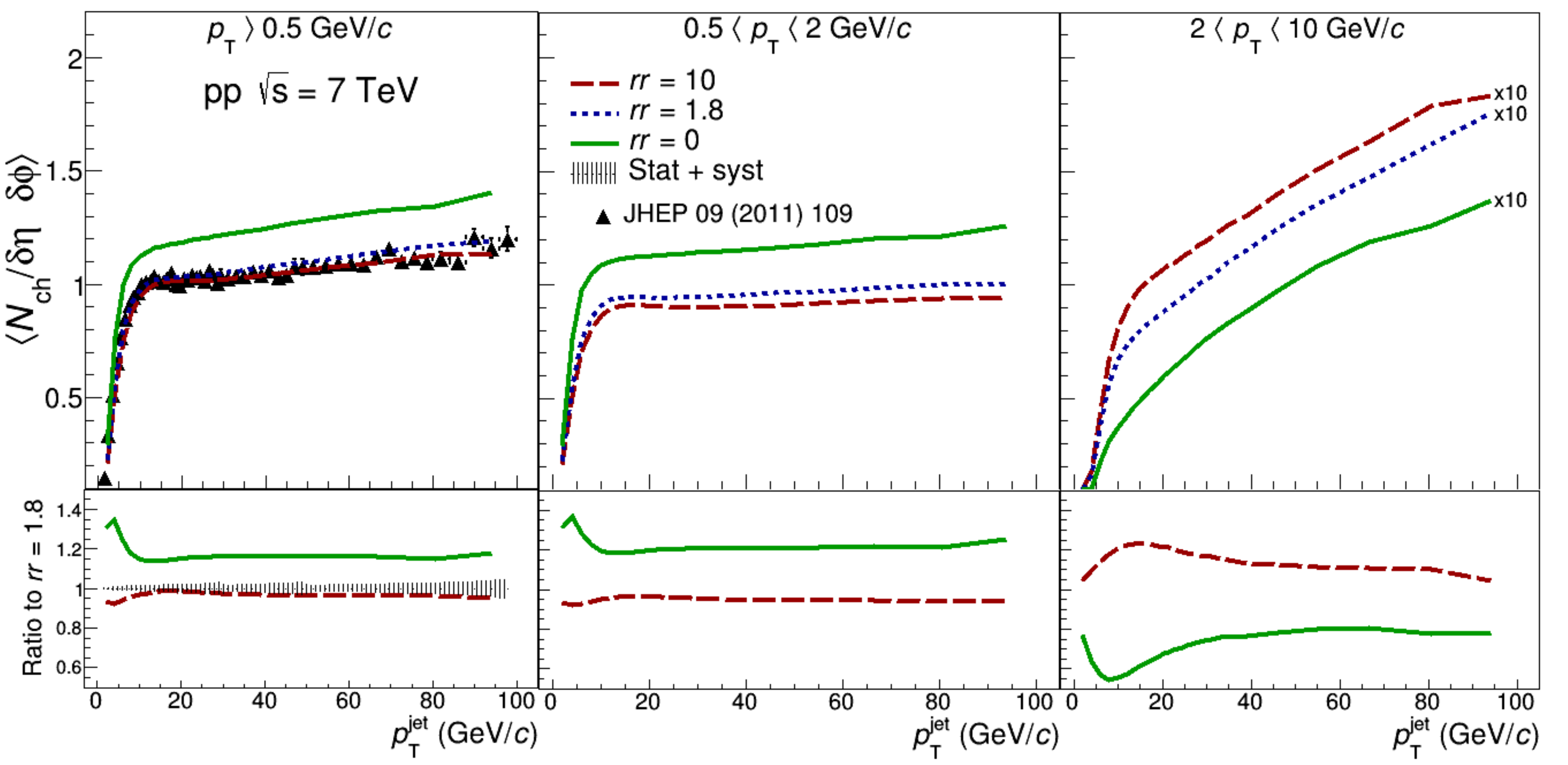}
\caption{ (Color online). The number density as a function of the $\ptjet$ for three different values of reconnection range. Results for the \pt-integrated case are displayed for both the CMS data and PYTHIA~8 simulations (left). Analogous results are shown for low- (middle) and intermediate- (right) \pt particles. Bottom panels show the variation of the number density with respect to $rr=1.8$. Shaded areas around unity indicate the statistical and systematic uncertainty (added in quadrature) on CMS data. }
\label{Ratios_nch}
\end{center}
\end{figure*}

Figure~\ref{Ratios_nch} shows the number density as a function of $\ptjet$ for $pp$ collisions at $\sqrt{s}=7$\,TeV. For the \pt-integrated case, CMS data are compared with PYTHIA~8 simulations considering three different values of $rr$. Within uncertainties, data are well described by simulations with $rr\geq1.8$. As expected, the number density decreases with increasing $rr$ due to the reduction of the number of color strings that connect the outgoing partons. Though, the \ptjet dependence of the number density is the same for the different $rr$ values. For the low-\pt case, we observe the same $rr$ ordering reported for the \pt-integrated case. However, the number density is reduced by $\sim 20$\% relative to the \pt-integrated case. The number density associated to intermediate-\pt particles  exhibits an opposite trend with increasing $rr$, i.e. going from low- to high-$rr$ values the number density increases suggesting a hardening of UE with increasing $rr$.

\begin{figure*}
\begin{center}
\includegraphics[keepaspectratio, width=1.0\columnwidth]{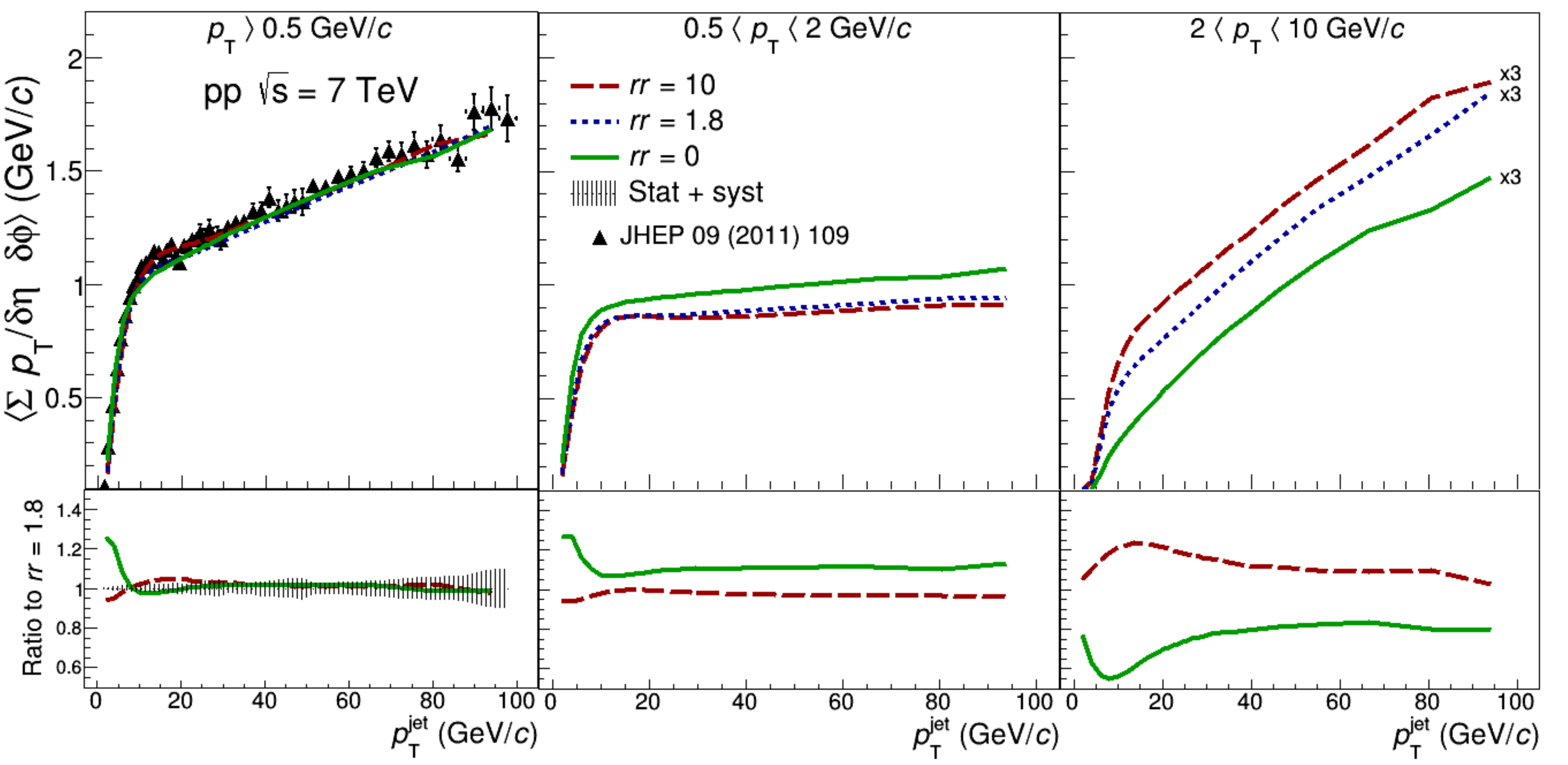}
\caption{ (Color online). The  summed-\pt density as a function of the $\ptjet$ for three different values of reconnection range. Results for the \pt-integrated case are displayed for both the CMS data and PYTHIA~8 simulations (left). Analogous results are shown for low- (middle) and intermediate- (right) \pt particles. Bottom panels show the variation of the summed-\pt density with respect to $rr=1.8$. Shaded areas around unity indicate the statistical and systematic uncertainty (added in quadrature) on CMS data.} 
\label{Ratios_pt}
\end{center}
\end{figure*}

The corresponding plots for the summed-\pt density are shown in Fig.~\ref{Ratios_pt}. Contrary to the number density, this observable presents a steeper rise with increasing \ptjet and no color-reconnection effects are visible for the \pt-integrated case. However, this does not happen when the summed-\pt density is calculated considering low- or intermediated-\pt charged particles. Namely, for low-\pt particles the summed-\pt density shows a saturation for \ptjet $>$ 10\,GeV/$c$. Moreover, the summed-\pt density decreases with increasing values of $rr$  (the CR effect amounts to 5\%). For the intermediate-\pt case, the summed-\pt density rises with increasing \ptjet more or less at the same rate that the \pt-integrated case.

%Both for Figure \ref{Ratios_nch} and \ref{Ratios_pt} the bottom panels show the variation of the UE observable with respect to $rr=1.8$. The ratios are performed relative to the default value of the reconnection range because, guided by the case of the number density, the data rules out $rr=0$. As a consequence, the effort is aimed to discriminate between $rr=1.8$ and $rr=10$.

In order to avoid any possible jet effect (at small values of $\lvert\Delta\eta\rvert$) that could bias the previous results, an extra cut on $\lvert \Delta\eta\rvert$ was tested. In this sense only particles with  $\lvert \Delta\eta\rvert > 1.6$ were selected, but no significant differences were observed.

In short, the behavior of the number density as a function of $\pt^{\rm jet}$ is determined by low-momenta particles while for summed-\pt density is determined by intermediate-\pt particles. It was verified that, qualitatively, the same behavior is obtained for both UE observables at $\sqrt{s}=900$\,GeV. It is worth mentioning that UE measurements considering different \pt intervals are not available. Our work suggests the necessity of such measurements in order to test the picture of PYTHIA~8, i.e. the hardening of the UE structure with increasing $\pt^{\rm jet}$ (or $\sqrt{s}$) by means of MPI and CR.

\paragraph{Flow-like effects within a hard UE environment}

It has been shown that CR produces flow-like effects that increase with increasing $\langle N_{\mathrm{MPI}} \rangle$~\cite{Ortiz:2013yxa}.  This is because the number of strings connecting scattered partons naturally increases with increasing  $\langle N_{\mathrm{MPI}} \rangle$. CR allows the color fields of the event to be redirected, relative to the case of color-separated MPI, in such a way that the total string length is reduced.  This produces a transverse boost to reconnected string pieces, that e.g. provides to heavier hadrons higher $\langle \pt \rangle$~\cite{Sjostrand:2018xcd}. It has been shown that in events with enhanced UE activity, relative to inelastic $pp$ collisions, the flow-like effects increase. This has been achieved by selecting highly-spherical events where no jet structure is present~\cite{Ortiz:2017jho}. Motivated by the similarity of the event-shape selection results with those for heavy-ion collisions where the proton-to-pion ratio (sensitive to radial flow~\cite{Adam:2015kca}) is studied for the regions sensitive to bulk production and jet fragmentation, separately. We decided to perform an analogous Monte Carlo analysis but considering the underlying event accompanying a very high-\pt jet. The goal of the study is to understand how the \ptjet selection affects observables sensitive to radial flow in the UE region.  

\begin{figure*}
\begin{center}
\includegraphics[keepaspectratio, width=0.9\columnwidth]{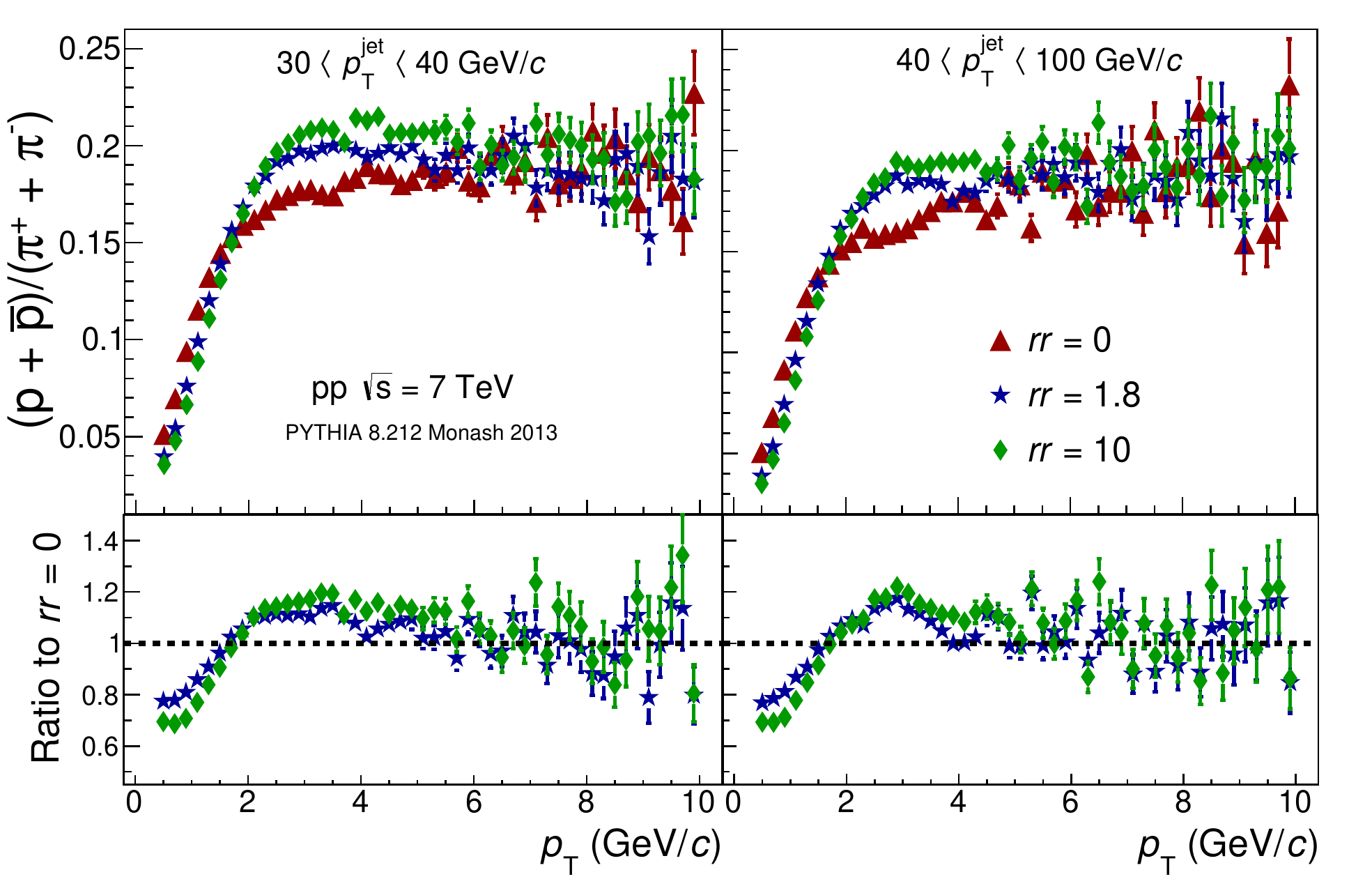}
\caption{ (Color online). Proton-to-pion ratio in the UE as a function of the hadron transverse momentum for different values of reconnection range in two \ptjet intervals: 30-40\,GeV/$c$ (left) and 40-100\,GeV/$c$ (right). Bottom panels show the variation of the proton-to-pion ratio with respect to $rr=0$. Error bars indicate the statistical uncertainties.} 
\label{flow-like}
\end{center}
\end{figure*}

Figure~\ref{flow-like} shows the proton-to-pion ratio as a function of the hadron transverse momentum in the underlying-event region for two \ptjet bins and for three different values of $rr$: 0, 1.8 and 10. The left-hand side of Figure~\ref{flow-like} shows the results for $30<\ptjet<40$\,GeV/$c$. Going from independent fragmentation ($rr=0$) to the largest value of $rr$, the proton yield is suppressed with respect to the pion one for $p_{\mathrm{T}}<2$\,GeV/$c$, whereas for intermediate transverse momentum the proton yield is enhanced relative to the pions. This is the mass effect attributed to color reconnection that was previously reported for the inclusive event~\cite{Ortiz:2013yxa}, i.e. without any separation of the soft and hard components. It is worth noting that the size of the bump shown in Fig.~\ref{flow-like} is smaller than that seen in the inclusive case~\cite{Ortiz:2013yxa}. The reduction is understood in terms of the hardness of the underlying event that increases with increasing \ptjet. The right-hand side of  Fig.~\ref{flow-like} shows the results when we move from $30<\ptjet<40$\,GeV/$c$ to higher jet transverse momentum ($40<\ptjet<100$\,GeV/$c$). Going from low- to high-\ptjet values the structure at $p_{\mathrm{T}}=3$\,GeV/$c$ is reduced. Moreover, the results show a smaller dependence on $rr$ indicating that the ratio is more sensitive to the fragmentation of hard partons. 

\begin{figure*}
\begin{center}
\includegraphics[keepaspectratio, width=0.9\columnwidth]{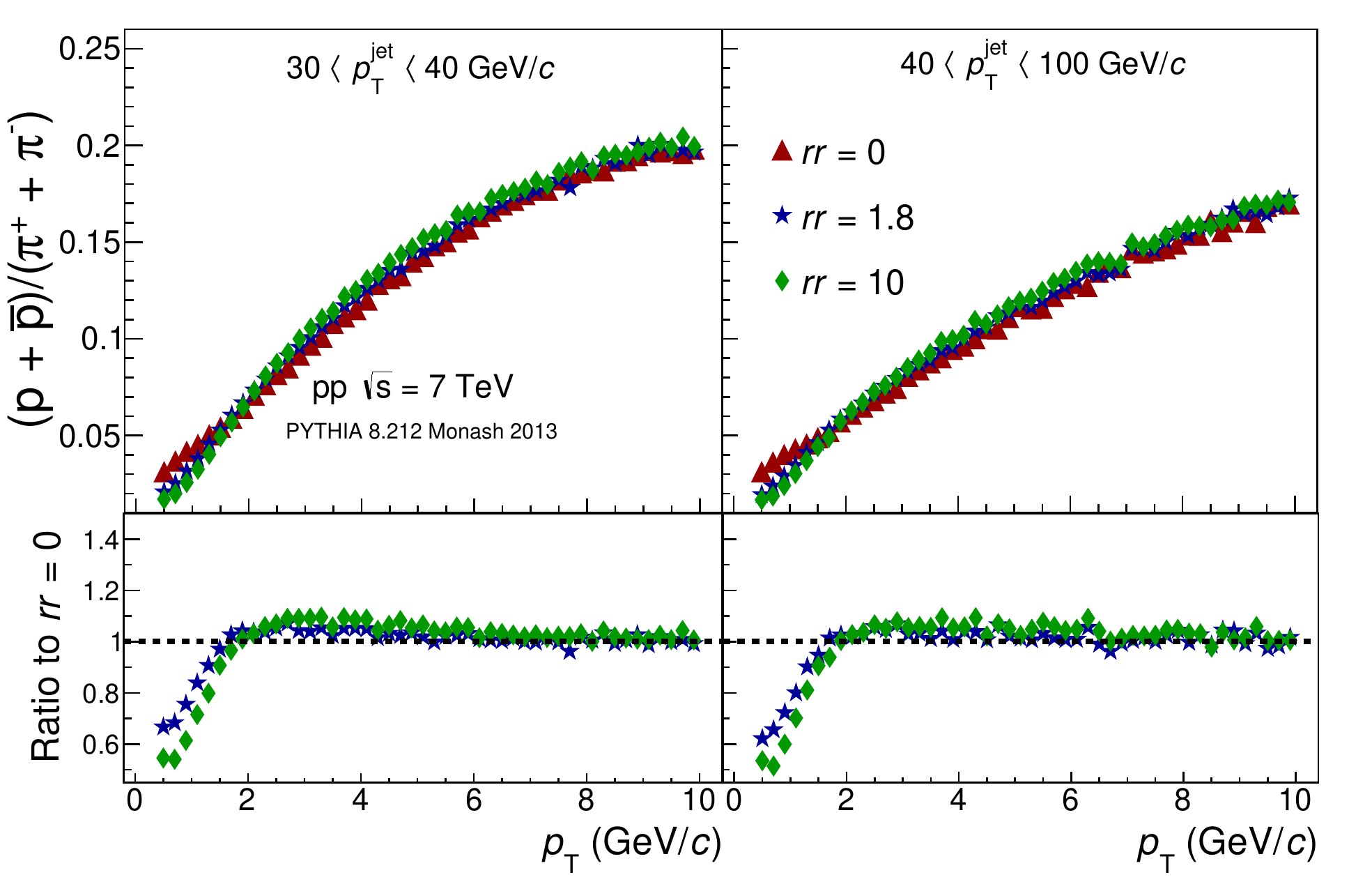}
\caption{ (Color online). Proton-to-pion ratio in the jet region and within $R<0.5$ as a function of the hadron transverse momentum for different values of reconnection range in two \ptjet intervals: 30-40\,GeV/$c$ (left) and 40-100\,GeV/$c$ (right). Bottom panels show the variation of the proton-to-pion ratio with respect to $rr=0$.  Error bars indicate the statistical uncertainties.} 
\label{noflow-like}
\end{center}
\end{figure*}

Finally, in order to show the distinct behavior of the UE with respect to jet fragmentation, Figure~\ref{noflow-like} presents the proton-to-pion ratio as a function of \pt obtained for the towards region ($\lvert \Delta \phi \rvert < \pi/3$). In particular, only pions and protons within $R<0.5$ are considered. Notice that, opposite to Figures~\ref{Ratios_nch} and~\ref{Ratios_pt}, the ratios in the bottom panels of Figures~\ref{flow-like} and~\ref{noflow-like} are performed relative to $rr=0$. The reason is because in the latter case the goal is to quantify the flow-like effect, if any, that is observed when color reconnection is present. 

Overall in Figure~\ref{noflow-like}, the ratio is significantly smaller than the corresponding for the underlying-event region. For example, while for the transverse region the ratio shows a maximum amounting to $\sim0.22$ at $\pt=3$\,GeV/$c$; for the jet peak  ($30<\ptjet<40$\,GeV/$c$) it reaches 0.1 and the value is subsequently reduced with increasing \ptjet. It is worth mentioning that although a color-reconnection effect is observed in the jet region as a consequence of the UE contamination, its effect is smaller than what is seen in the transverse side. The effects observed for the UE and jet regions are qualitatively similar to those measured in heavy-ion collisions resulting from the jet-bulk separation using leading-hadron correlations. The statement there~\cite{Veldhoen:2012ge} is that the flow-like peak observed in proton-to-pion ratio for the most central Pb-Pb collisions at the LHC energies is a medium effect.

The effects reported in this paper should be searched in data; the outcome would contribute to the understanding of the unexpected heavy ion-like signals discovered in small systems by LHC experiments~\cite{Khachatryan:2016txc,ALICE:2017jyt,Acharya:2018orn}.

\section{Conclusion}

Motivated by the results of our previous work~\cite{Ortiz:2017jaz}, where we shown that the particle production sensitive to the underlying event (UE observables) exhibited an approximate KNO scaling. In this paper we investigated in more detail the scaling properties using data from three experiments at the LHC, as well as their description using multiple partonic interactions and color reconnection in the context of PYTHIA~8 (version 8.212 tune Monash 2013) event generator.

We studied the number density in the transverse side divided by the charged-particle density in inelastic $pp$ collisions (KNO variable). Both two quantities were calculated considering charged-particles with $\pt>0.5$\,GeV/$c$. The ratio was found to increase with increasing $\sqrt{s}$, the trend was well described by PYTHIA~8. In particular going from $pp$ collisions at $\sqrt{s}=0.9$ to 7 TeV, the ratio increased by 8-10\%.  In PYTHIA~8 the rate of such an increase with increasing $\sqrt{s}$ was close to that for multiple partonic interactions. Moreover, using the limited available data we found indications that particle production at intermediate-\pt ($2<\pt<10$\,GeV/$c$) increased faster in UE than in inelastic events. Whereas at lower \pt, such an increase in both UE and inelastic events was the same within uncertainties. This effect could yield the small breaking (10\%) of the KNO scaling in the particle production sensitive to UE.  

The results above motivated the study of the traditional underlying-event observables considering particles within different \pt intervals. We found that color reconnection modifies the \pt spectra of unidentified charged particles, producing a variation in the number and summed-\pt densities. In short, color reconnection enhanced the particle production at intermediate transverse momenta in regions far from the jet peaks. The modification was mass dependent, e.g. the proton-to-pion ratio as a function of \pt in the underlying-event region exhibited a flow-like response with the variation of the color reconnection strength. This behavior was very similar to what was observed in the bulk (everything outside the jet peak) region measured in heavy-ion collisions. 

Our paper encourages the measurement of \pt spectra of unidentified and identified charged particles in the transverse side. The variation of such \pt spectra as a function of center-of-mass energy, \ptjet (or $\pt^{\rm leading}$), and system size would add key information on the origin of the heavy ion-like effects observed in $pp$ collisions.
 
\section{Acknowledgments}
We acknowledge the technical support of Luciano Diaz and Eduardo Murrieta for the maintenance and operation of the computing farm at ICN-UNAM. Support for this work has been received from CONACyT under the Grant No. 280362 and PAPIIT-UNAM under Project No. IN102118.
%\section*{References}

\bibliography{mybibfile}

\end{document}